\documentclass[preprint,12pt]{elsarticle}

\usepackage{graphicx}
\usepackage{subcaption}
\usepackage{xcolor}
\usepackage{amssymb}
\usepackage{amsmath}
\usepackage{float}
\usepackage{algorithm}
\usepackage[noend]{algpseudocode}
\usepackage{hyperref}
\usepackage[utf8]{inputenc}

\newcommand{\eref}[1]{\eqref{#1}}
\newcommand{\ee}{\mathbb{E}}

\newcommand{\pf}{\pi}
\newcommand{\pred}{p}
\newcommand{\act}{a}
\newcommand{\state}{s}
\newcommand{\reward}{\mathrm{rwd}}

\newcommand{\ret}{r}
\newcommand{\snoise}{\eta}
\newcommand{\rnoise}{\xi}
\newcommand{\pnoise}{\eta^{(p)}}
\newcommand{\anoise}{\eta^{(a)}}
\newcommand{\retnoise}{\eta^{(r)}}
\newcommand{\qf}{\mathcal{Q}}
\newcommand{\vf}{\mathcal{V}}
\newcommand{\policy}{\phi}
\newcommand{\params}{\Theta}

\journal{The Journal of Finance and Data Science}

\begin{document}

\begin{frontmatter}


\title{Deep Deterministic Portfolio Optimization}


\author[cfm,central]{Ayman Chaouki}
\author[cfm]{Stephen Hardiman}
\author[cfm,x]{Christian Schmidt}
\author[cfm]{Emmanuel S\'eri\'e}
\author[cfm]{Joachim de Lataillade}
\address[cfm]{Capital Fund Management, 23 rue de l'Universit\'e, 75007, Paris}
\address[x]{Chair of Econophysics and Complex Systems, \'Ecole Polytechnique, Palaiseau}
\address[central]{\'Ecole Centrale-Supélec, Gif-Sur-Yvette}

\begin{abstract}
  Can deep reinforcement learning algorithms be exploited as solvers for optimal trading
  strategies?  The aim of this work is to test reinforcement learning algorithms on
  conceptually simple, but mathematically non-trivial, trading environments. The
  environments are chosen such that an optimal or close-to-optimal trading strategy is
  known.  We study the deep deterministic policy gradient algorithm and show that such a
  reinforcement learning agent can successfully recover the essential features of the
  optimal trading strategies and achieve close-to-optimal rewards.
\end{abstract}


\end{frontmatter}

\section{Introduction}

The fusion of reinforcement learning (RL) with deep learning techniques, aka.\ deep RL
(dRL), has experienced an astonishing increase in popularity over the last years
\cite{henderson2018matters}.  Undoubtedly, dRL has been successfully applied to a broad
range of applications with major success ranging from playing games, controlling robots, trading and even solving complex physics problems
\cite{silver2017go, mnih2015atari, moravvcik2017deepstack, brown2017libratus,
  deng2016deeptrading, pan2019autoregressive, levine2016robo, pan2017drive,
  franccois2016energy}.

Finance has been among the many branches that devoted much attention to reformulate their
problems in a dRL framework to find new algorithmic avenues to solve them
\cite{ritter2017,dutting2017auctions}. One subject of interest has been the application of
(d)RL to dynamical portfolio allocation \cite{ritter2017, deng2016deeptrading,
  moody2001trade, yu2019model_based, neuneier1996optimal}.  
  
Besides the astonishing practical results, dRL has been criticized to possess a reproducibility issue
\cite{henderson2018matters, recht2019tour, hutson2018artificial}.  And indeed, to the best
of our knowledge, the dRL algorithms for portfolio allocation that can be found in the
literature are typically only compared with respect to sub-optimal reference strategies
(with some restricted exceptions, such as \cite{neuneier1996optimal}).

Little effort has been devoted to the question of whether dRL can learn known optimal
trading strategies for the dynamical portfolio allocation problem.  In this note we
address this question and explore the potentials and pitfalls of dRL applied to portfolio
allocation.  We do so by testing the widely used deep deterministic policy gradients
(DDPG) algorithm against an ensemble of three different trading problems for which the
\emph{optimal (or close-to-optimal) control strategies} are known.

\subsection{Dynamic Portfolio Optimization and reinforcement learning}

Formally, the RL problem is a (stochastic) control problem of the following form:
\begin{align}
	\begin{split}
    \max_{\{\act_t\}} & \  \ee\left[ \sum_{t=0}^{T-1} \reward_t(\state_t, \act_t, \state_{t+1}, \rnoise_t)\right]
    \\
    \mathrm{s.t.} & \ \state_{t+1} = f_t(\state_t, \act_t, \snoise_t)
    \, ,
    \end{split}
    \label{eq:rl_problem} 
\end{align}
where $\act_t \in \mathcal{A}$ indicates the control, aka.\ actions,
$\state_t \in \mathcal{S}$ the state of the system at time $t$, $\snoise_t$ and $\rnoise_t$
are noise variables, and $\reward_t$ is the reward received at every time step.  In
RL the second line in the above equation is usually referred to as the `environment'.  The
`agent' intends to choose its actions $\act_t$, given the state $\state_t$, so as to
maximize the total expected accumulated reward.

Comparing this with the setting of dynamic portfolio optimization reveals the link between
the two problems.  For the sake of simplicity we will restrict the following discussion to
the one dimensional case.  In dynamic portfolio optimization the investor aims to
dynamically allocate his weight $\pf_t \in \mathbb{R}$ on an asset, that yields returns
$\ret_t$, such that the expected utility $U(\cdot)$ of future wealth is maximized:
\begin{eqnarray}
	\begin{split}
    \max_{\{\pf_t\}} & \  \ee\left[ U(\sum_{t=0}^{T-1} \mathrm{PnL}_{t, t+1} )\right]
    \\
    \mathrm{s.t.} & \ \mathrm{constraints}
    \, . 
    \end{split}
    \label{eq:po_problem}
\end{eqnarray}
where $\mathrm{PnL}_{t,t+1}$ stands for `profit and loss', i.e.
$\mathrm{PnL}_{t,t+1} = \mathrm{gain}_{t+1} - \mathrm{cost}_{t,t+1}$, with
$\mathrm{gain}_t = \pf_{t} \ret_t$ the profit that the portfolio yields at every time step
and $\mathrm{cost}_{t,t+1}$ some cost function that
incorporates trading cost and/or other fees and effective costs.

In general this latter formulation \eref{eq:po_problem} encompasses a much bigger class of
problems (and in particular the former \eref{eq:rl_problem}).  In this work we are
interested in problems that lay at the intersection of \eref{eq:rl_problem} and \eref{eq:po_problem}.  In particular we shall consider problems for which the utility
maximization can be brought into the following form:
\begin{eqnarray}
    \max_{\{\pf_t\}} 
    \ee\left[
    \sum_{t=0}^{T-1} \pf_{t+1} \ret_{t+1} - \mathrm{cost}(|\pf_{t+1}-\pf_t|) - \mathrm{risk}(\pf_{t+1})
    \right]
    \, .
\end{eqnarray}
The risk term originates from the shape of the utility function and can be thought of as a
regularizer that penalizes risky portfolio weights. For example
$\mathrm{risk}(\pf_{t}) = \pf_{t}^2$ punishes large positions for being more risky.  One
important class of problems that fall into this framework are the `mean-variance
equivalent' problems \cite{markowitz1952portfolio, chamberlain1983meanvariance}.

In order to bring the above problem into the RL framework it is natural to define the
actions as the trades between subsequent time steps $\act_t = \pf_{t+1} - \pf_t$.  It is
less obvious what variables should play the role of the state $\state_t$ according to
which the agent takes its action.  A standard assumption is that the returns $\ret_t$ can
be decomposed into a predictable and an unpredictable noise term, $\pred_t$ and
$\retnoise_t$ respectively:
\begin{equation}
    \ret_{t+1} = \pred_t + \retnoise_t
    \, .
\end{equation}
The dynamic portfolio optimization can now be recast as an RL problem:
\begin{eqnarray}
	\begin{split}
    \max_{\{\act_t\}}\ & \ee \left[ \sum_{t=0}^{T-1} \pf_{t+1} \ret_{t+1} - \mathrm{cost}(|\act_t|) - \mathrm{risk}(\pf_{t+1})\right]
    \\
    \mathrm{s.t.}\ & 
    \begin{cases}
        \pf_{t+1} &= \pf_t + \act_t
        \\
        \pred_{t+1} &= f(\pred_t) + \pnoise_t
        \\
        \ret_{t+1} &= \pred_t + \retnoise_t
    \end{cases}
    \, . 
    \end{split}
    \label{eq:dpo_rl_problem}
\end{eqnarray}
Where we additionally assumed that $\pred_t$ follows a Markovian dynamic.

Most control problems of the above form are not known to have closed form solutions and
require some sort of algorithmic approach.  The need for RL comes to bear, when the
problem \eref{eq:dpo_rl_problem} has difficult reward or state transition functions that
result in non-linear control policies.  The \emph{deep} RL framework becomes particularly
necessary when the control and/or action spaces are continuous and traditional methods,
such as Q-learning, fail.  It is most challenging in the model-free context where no
further model assumptions are made (neither on the reward function, nor on the dynamical
equations of the state) and the only source of information are the accumulated rewards
from which the trading policy and/or value function must be deduced. This will be the
setting considered in this manuscript.

\subsection{Contributions}

Applying RL to portfolio optimization is no new idea \cite{neuneier1996optimal,
  moody2001trade}, indeed it is almost as old as the model-free RL algorithms
\cite{watkins1989qlearning, sutton2018reinforcement}.  However, to the best of our
knowledge, besides several publications on modern dRL strategies for trading, these
attempts never seem to have been systematically evaluated against known optimal
strategies.

Our contribution is to propose an ensemble of dynamic portfolio optimization problems for
which optimal (or close-to-optimal) control policies are known and to test DDPG against
these strategies. This reveals the potential and pitfalls of current dRL approaches to
this problem.  The set of problems that we propose are also of interest beyond dynamic
portfolio optimization as they provide testing environments with conceptually simple, but
mathematically challenging control policies.  The environments and code is accessible
through our repository\footnote{\url{https://github.com/CFMTech/Deep-RL-for-Portfolio-Optimization}}.

\subsection{Related work}

Modern portfolio optimization has a long history that goes back to Markowitz
\cite{markowitz1952portfolio} and Merton \cite{merton1969lifetime}. In these initial
formulations trading costs were typically neglected.  
If, however, cost is present the optimal strategy must (a) plan ahead to take into account possible auto-correlation in the predictor and (b) trade lightly to take into account the cost \cite{lataillade2012optimal, abeille2016lqg, bertsimas1998optimal,
  garleanu2013dynamic, martin2011mean, muhle2017primer}.
  
There are special cases of the
problem \eref{eq:dpo_rl_problem} that can be solve in a closed form
\cite{garleanu2013dynamic, lataillade2012optimal, martin2011mean} and that we will take as
reference solutions. Further details will be discussed in section \ref{sec:setting_stage}.
In general, however, closed form solutions of such stochastic control problems are scarce
and it is necessary to resort to approximations.

Traditional methods are often (but not
always) based on dynamic programming \cite{bertsekas1995dynamic, garleanu2013dynamic,
  bertsimas1998optimal}, model-predictive control and convex optimization,
cf. \cite{boyd2017} and references therein.  Model-free reinforcement learning is an
alternative approach that does not assume a model of the system and takes decision solely
from the information received at every time step through the rewards in
\eref{eq:dpo_rl_problem}.  Early works that are applying this idea to dynamic portfolio
allocation can be found in \cite{neuneier1996optimal, bertsimas1998optimal,
  moody1999reinforcement, moody2001trade}. However, these approaches were mostly limited
to low-dimensional, discrete state and action spaces.  The rise of deep techniques has
revived the interest in applying dRL strategies to more complicated and/or continuous
settings \cite{franccois2018drlreview} and more modern (d)RL approaches for dynamical
portfolio allocation can e.g. be found in \cite{deng2016deeptrading, yu2019model_based}.

A somewhat related approach to what we are are doing was followed by the authors of
\cite{mehta2018} that investigated if dRL strategies successfully learn known optimal
algorithmic strategies for online optimization problems.

\section{Setting the stage}
\label{sec:setting_stage}

\subsection{The environments and their reference solutions}

This section establishes the three environments that the algorithm is tested against.  We
start by introducing the state space, i.e.\ the dynamics $\state_{t} \mapsto \state_{t+1}$
and then define the three different reward functions \eref{eq:dpo_rl_problem}, each of
which defines a different environment with distinct optimal reference solutions, i.e.\ controls.

Throughout the manuscript we assume that the predictor $\pred_t$ is an autoregressive (AR)
process with parameter $\rho$, normalized to unity equilibrium variance.  The variables
$\pf_t, \pred_t, \ret_t$ thus evolve according to
\begin{align} 
    \pf_{t+1} &= \pf_t + \act_t 
    \label{eq:pf_state_space}
    \\
    \pred_{t+1} &= \rho \pred_t + \pnoise_t
    \label{eq:pred_state_space}
    \\
    \ret_{t+1} &= \pred_t + \retnoise_t 
    \label{eq:ret_state_space}
    \, .
\end{align}
Note that the agent's state $\state_t = (\pf_t, \pred_r)$ only contains the observables
available at time $t$.

It is important to note that \eref{eq:dpo_rl_problem} together with
\eref{eq:pf_state_space}--\eref{eq:ret_state_space} in principle permits to eliminate the
returns $\{\ret_t\}$ entirely; simply by replacing $\ret_{t+1} \mapsto \pred_t$ in the
rewards.  In this case, the agent has perfect state observations in that the reward
received at time $t$ is entirely composed of observables accessible to the agent:
$\pf_t, \pred_t$ and $\act_t$.  In practice, however, this is an unrealistic assumption
and one should rather work with the rewards that contain the additional noise from the
returns.

In the first case one simply asks if RL can find back (almost) optimal stochastic control
strategies \emph{under perfect state information}. In the second case the problem becomes
a bit harder, as the rewards contain \emph{additional noise} due to unobserved variables.
We will start our experiments by considering the simpler former case and then add the
additional noise and compare.

The reference solutions that will be derived below are valid in
both these cases.

\subsubsection{Environment with quadratic cost and quadratic risk control}

If the cost- and risk-terms are quadratic and the above dynamics
\eref{eq:pf_state_space}--\eref{eq:ret_state_space} hold, the problem
\eref{eq:dpo_rl_problem} becomes a finite-horizon, discret-time linear–quadratic
regulator (LQR).
\begin{eqnarray}
	\begin{split}
    \max_{\{\act_t\}}\ 
    &
    \ee\left[ \sum_{t=0}^{T-1} \pf_{t+1} \pred_t - \Gamma a_t^2  -  \lambda \pf_{t+1}^2  \right] 
    \\ 
    \mathrm{s.t.}\ 
    & 
    \eref{eq:pf_state_space} \textrm{ and } \eref{eq:pred_state_space}
    \, .
    \end{split}
    \label{eq:lqr_base}
\end{eqnarray}
When considering the average expected gain within an infinite horizon as described in
\cite{bertsekas1995dynamic} (i.e.\ dividing the global gain by $T$ and taking the
limit $T\to \infty$), one can derive a closed-form solution for the optimal control:
\begin{equation}
\act_t = -K \state_t  \, ,
\end{equation}
where $K$ follows from the discrete Riccati-equations and which has a linear dependence on the state.

Without cost term, the solution would be to adapt the portfolio at every time step
according to the Markowitz-allocation
\begin{equation}
    \pf_{t+1}^* = \pf_{t+1}^{\mathrm{(M)}} = \frac{1}{2 \lambda} \pred_t 
    \label{eq:markowitz}
    \, .
\end{equation}
The cost-term adds friction and slows down the trading strategy as compared to the
Markowitz allocation and the optimal solution for $t+1$ is a damped version of the
Markowitz allocation \eref{eq:markowitz} with the current portfolio
\cite{garleanu2013dynamic}. Equivalently the optimal portfolio can be expressed as an
exponential moving average of the predictors at $t,t-1,\dots$.
\begin{align}
    & \pf_{t+1}^*
    = 
    (1-\omega) \pi_t + \omega \psi \pf_{t+1}^{\mathrm{(M)}}
    \, ,
\end{align}
with $\psi=\frac{\omega}{1-(1-\omega)\rho}$,
$\omega=f_{\mathrm c}\left(\sqrt{\frac{\lambda}{\Gamma}}\right)$ and
$f_{\mathrm c}(x) = \frac{2}{1 + \sqrt{1 + \frac{4}{x}}} = \frac{x}{2}(\sqrt{x^2 + 4} -
x)$. Note that both $\omega$ and $\psi$ have values between $0$ and $1$.

In all reported experiments with this environment, we used $\Gamma = 1$, $\lambda=0.3$ and
$\rho=0.9$.

\subsubsection{Environment with proportional cost and quadratic risk}

Another interesting setting is the case of proportional (or linear) costs:
\begin{eqnarray}
	\begin{split}
    \max_{\{a_t\}} 
    &\ 
    \mathbb{E}\left[
    \sum_{t=0}^{T-1}
    \pf^\intercal_{t+1} \pred_t - \Gamma |\act_t|  -  \lambda \pf_{t+1}^2  
    \right] 
    \\ 
    \mathrm{s.t.} 
    &\ 
    \eref{eq:pf_state_space} \textrm{ and } \eref{eq:pred_state_space}
    \, .
    \end{split}
    \label{eq:band_problem}
\end{eqnarray}
For this problem, the optimal stochastic control strategy can be derived
\cite{martin2011mean} and takes the form of a no-trading band system around the (rescaled)
predictor's value:
\begin{equation}
    \pf^*_{t+1}
    =
    \begin{cases} 
    u(\pf_{t+1}^{\mathrm{(M)}}) & \textrm{if}\ \pf_{t} > u(\pf_{t+1}^{\mathrm{(M)}}) 
    \\ 
    l(\pf_{t+1}^{\mathrm{(M)}}) & \textrm{if}\ \pf_{t} < l(\pf_{t+1}^{\mathrm{(M)}})
    \\ 
    \pf_t & \textrm{otherwise} 
    \end{cases}
\end{equation}
The edges $u(\cdot)$ and $l(\cdot)$ are rather non-trivial, but it is a good approximation
\cite{martin2011mean,lataillade2012optimal,muhle2017primer} to consider the band as being
symmetric around the (rescaled) predictor and of constant size:
\begin{equation}
    u(\pf_{t}^{\mathrm{(M)}})=\pf_{t}^{\mathrm{(M)}}+b\ \ \textrm{and}\ \ l(\pf_{t}^{\mathrm{(M)}})=\pf_{t}^{\mathrm{(M)}}-b
\end{equation}
In this paper we will use this heuristic as our benchmark, with the parameter $b$ simply
given by a numerical grid search.

In all reported experiments with this environment, we used $\Gamma = 4$, $\lambda=0.3$ and $\rho=0.9$.

\subsubsection{Environment with proportional cost and maxpos risk}

Finally, it is also meaningful to consider other forms of risk constraints. In order to
test this, we consider the `maxpos' constraint which imposes a maximum
constraint on the absolute positions:
$|\pf_t| \leq M$. Formally, that is:
\begin{eqnarray}
    \max_{\{|\pf_t| \leq M\}} 
    &
    \ee\left[
    \sum_{t=0}^{T-1}
    \pf_{t+1} \pred_t - \Gamma |\act_t| 
    \right]
    \\
    \mathrm{s.t.}
    &
    \eref{eq:pf_state_space} \textrm{ and } \eref{eq:pred_state_space}
    \, .
    \label{eq:bangbang_problem} 
\end{eqnarray}
The optimal trading strategy is a threshold-based controller that trades to the permitted
maximum position whenever the predictor overcomes a value $q$
\cite{lataillade2012optimal}, i.e.
\begin{equation}
    \pf^*_{t+1}
    =
    \begin{cases} 
        M & \textrm{if}\ \pred_t > q 
        \\
        -M & \textrm{if}\ \pred_t < -q
        \\
        \pf_t & \textrm{otherwise}
    \end{cases}
    \, .
\end{equation}
Again, the subtlety is to find the right value of $q$, which is derived in~\cite{lataillade2012optimal}. Here again, we will use a
brute-force grid search to find a good approximation of this threshold.

In all reported experiments with this environment, we used $\Gamma = 4$, $M = 2$ and
$\rho = 0.9$.

\subsection{The algorithm}

We selected Deep Deterministic Policy Gradient (DDPG) \cite{lillicrap2015continuous} for
the task. It is simple, yet state-of-the-art in continuous control problems and further
has been employed previously \cite{xiong2018ddpgtrading}.

For convenience of the reader we recap the main elements of the DDPG algorithm in
\ref{sec:appendix_algo_details} and outlined the details we employed for successful
training. A summary of the DDPG algorithm used in this work is given in the pseudo-code
below in Alg.~\ref{alg:our_ddpg}, and an implementation is available through our
repository.

\begin{algorithm}
\caption{DDPG with PER}
\begin{algorithmic}[1]
\State Define the predictor AR process
\State Initialize networks $\qf^w$ and $\policy^\params$, and the replay buffer of fixed size
\State Initialize target networks $\widetilde{\qf}^{\widetilde{w}}$ and $\widetilde{\policy}^{\widetilde{\params}}$
\State Initialize the environment
\State Initialize the exploration noises $\left( \anoise_t\right)_{1 \leq t \leq  T_{pretrain}}$
\For{$t=1$ to $T_{pretrain}$}
\State Observe state $\state_t$
\State Take action $\act_t = \policy^\params\left( s_t\right) + \anoise_t$
\State Observe reward $\reward_t$ and next state $\state_{t+1}$
\State Add $\left(\state_r, \act_t, \reward_t, \state_{t+1} \right)$ with priority $p=|\reward_t|$ to the replay buffer
\EndFor
\For{$\mathrm{episode}=1$ to $n$}
\State Initialize the environment
\State Initialize the exploration noises $\left( \anoise_t\right)_{1 \leq t \leq  T}$.
\For{$t=1$ to $T$}
\State Observe state $\state_t$
\State Take action $\act_t = \policy\left( s_t\right) + \anoise_t$
\State Observe reward $\reward_t$ and next state $\state_{t+1}$
\State Add $\left(\state_r, \act_t, \reward_t, \state_{t+1} \right)$ with the highest priority in the buffer
\If{$t \equiv 0 \hspace{1mm} (\textrm{mod}\ \tau)$}
\State Sample a batch $\left(\state_{t_i}, \act_{t_i}, \reward_{t_i}, \state_{t_i+1} \right)_{1 \le i \le b}$ with prioritized sampling according to probabilities $P\left( i\right) = \frac{p_i^\alpha}{\sum_k p_k^\alpha}$ with $0 \le \alpha \le 1$:
\State Compute $\forall{i}\in \{1, \ldots, b \}: \widetilde{Q}_i = \reward_{t_i} + \gamma \widetilde{\qf}\left(\state_{t_i+1}, \widetilde{\policy}\left(\state_{t_i+1} \right) \right)$
\State Compute $\forall{i}\in \{1, \ldots, b \}: \delta_i = |\qf\left(s_i, a_i \right) - \widetilde{Q}_i|$
\State Compute weights $\forall{i}\in \{1, \ldots, b \}: \alpha_i = \left(\frac{1}{N}\frac{1}{P\left(i \right)} \right)^\beta$
\State Normalize weights $\forall{i}\in \{1, \ldots, b \}: \alpha_i \leftarrow \frac{\alpha_i}{\max_k\left( \alpha_k\right)}$
\State Update priorities $\forall{i}\in \{1, \ldots, b \}: p_i = |\delta_i|$
\State Gradient update : $\mathcal{L}_{critic}\left( w\right) = \frac{1}{b}\sum_{i=1}^b \alpha_i\delta_i^2$
\State Gradient update : $\mathcal{L}_{actor}\left(\params \right) =
-\frac{1}{b}\sum_{i=1}^b\qf\left(\state_{t_i}, \policy^\params\left( \state_{t_i}\right)
\right)$ 
\State Update target networks \begin{align*}
    \Tilde{w} &\leftarrow \tau_{critic}w + \left(1 - \tau_{critic} \right)\Tilde{w} \\
    \widetilde{\params} &\leftarrow \tau_{actor}\params + \left(1 - \tau_{actor} \right)\widetilde{\params}
\end{align*}
\EndIf
\EndFor
\EndFor
\end{algorithmic}
\label{alg:our_ddpg}
\end{algorithm}

The main ingredients that we added where (a) a prioritized replay buffer and (b) an
additional soft thresholding function in the cost-term for the environment with maxpos
risk control.

\begin{figure}[h!]
    \centering
    \includegraphics[width=0.5\linewidth]{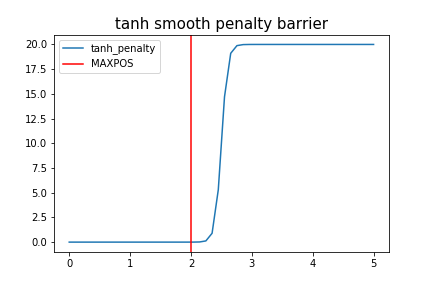}
    \caption{\textrm{tanh} barrier penalty used to stabilize training}
    \label{fig:tanh_barrier}
\end{figure}

Since the maxpos risk constraint is not differentiable we need an additional
trick to adapt DDPG to enable better convergence. The first ingredient is to simply clip the positions
resulting from the actor network's actions such that the constraint is not violated.  However, since this leads to poor gradient
estimations, we add an additional cost to the reward function, using a smooth penalty for
positions with magnitude beyond the maxpos, as depicted on
Figure~\ref{fig:tanh_barrier}. The reward function becomes:
\begin{align*}
  \reward\left(\pred, \pf, \act \right) = \pred \pf - \psi|\act | - \beta \left\{\textrm{tanh}\left[ \alpha (|\pf+\act| - (1+\gamma) M) \right] + 1\right\}
  \, .
\end{align*}

In our experiments we chose $\beta=10, \alpha=10, \gamma=\frac{1}{4}$. The addition of the
$\tanh$-barrier is crucial to stabilize training. Without it, out of the 16 agents we
trained, half diverged even under perfect state observations. The penalty prevents this divergence (at
least under perfect state observability). The use of $\tanh$ is justified by its smooth and upper
bounded nature, we tested a constant penalty and an exponential one with no success.


\section{Results}

\begin{figure}[ht]
	\centering
	\includegraphics[width=\textwidth]{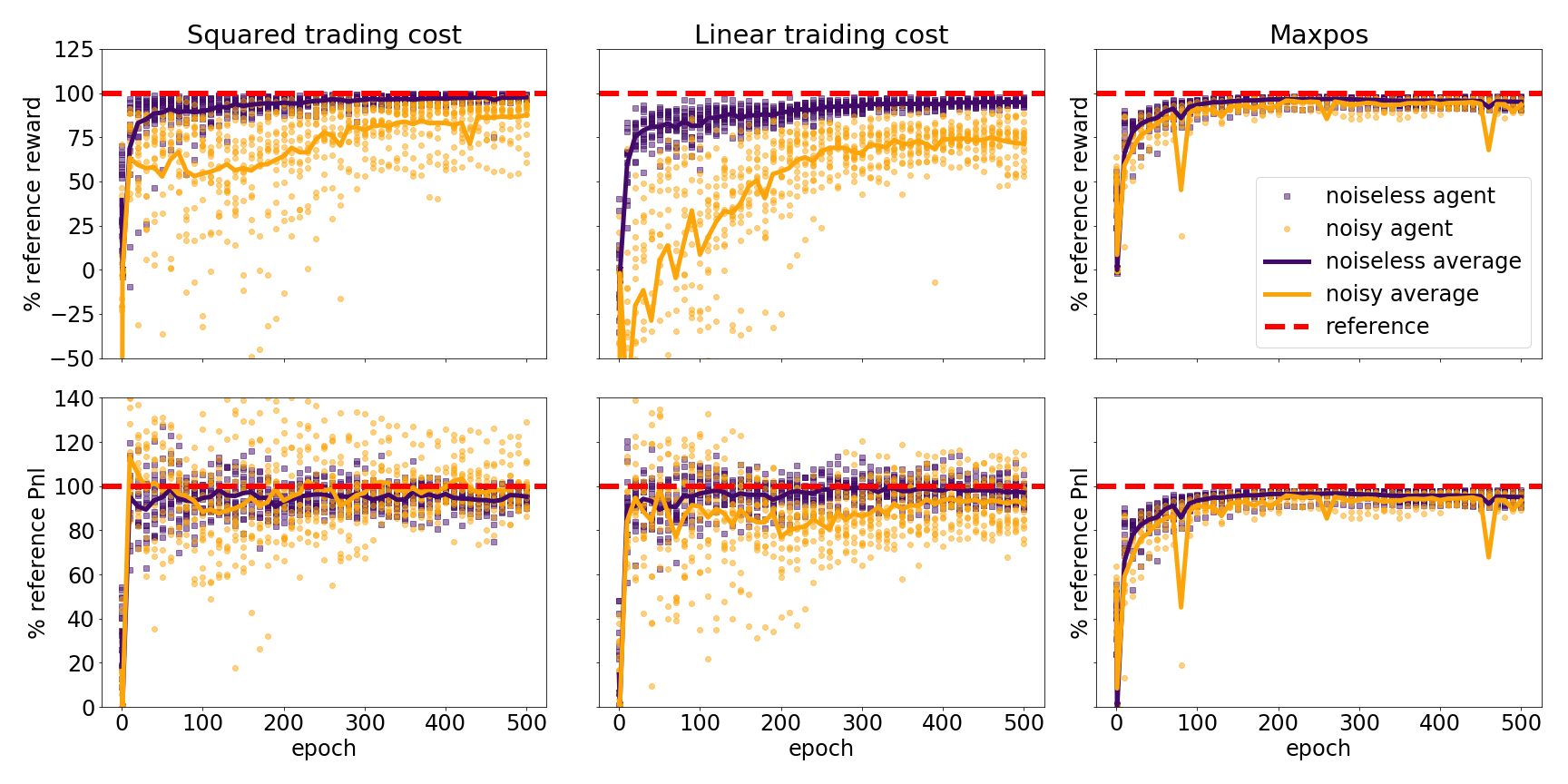}
	\caption{Results for the setting where the agent has perfect state information
          (purple markets) and under additional noise in the returns (yellow markers).  In
          both settings we report the theoretical rewards and PnLs under perfect state
          information. The red dashed horizontal line represents the reference agents from section 
          \ref{sec:setting_stage} and we plot the achieved relative performance difference
          w.r.t.\ the reference.  For every reported epoch each point represents one
          agent, i.e.\ a different random seed for the learning. The performances are
          measured as the mean over ten out-of-sample environments, each $T=5000$ time
          steps long.  The solid colored lines represent the average over the $16$
          different random seeds, i.e.~the average among agents.  For the `maxpos'
          environment we report only those agents that converged, hence there are fewer
          points and the figure is strongly biased towards the positive results.}
	\label{fig:training}
\end{figure}

We start by reporting the results of the training of the algorithm for all three
environments in Fig.~\ref{fig:training}.  The following conclusions can be drawn:
\begin{itemize}
\item For all three environments the algorithm successfully learns trading strategies with
  close-to-optimal rewards and PnLs. See also the tables for further quantitative
  comparison.
\item The PnL achieves close-to-optimal performance faster than the rewards, hinting at
  the fact that the algorithm first learns to follow the signal and to control the cost
  before it fine-tunes the risk-control. This is also represented in the wider spread of
  the points in the early stages of training and more concentration in the end.
\item The most challenging environment is the third environment, with linear cost and
  maxpos risk control for which not even all agents converged.
\item The environments for which additional noise is present in the rewards are harder:
  lower overall performance is achieved and there is more variability in the
  results.\footnote{The noise-to-signal ratio applied during the training was
    $\sigma(\retnoise_t) = 10$ for the environments with quadratic risk penalty and
    $\sigma(\retnoise_t) = 4$ for the one with maxpos risk.}
\end{itemize}

\begin{figure}
    \centering
    \includegraphics[width=\textwidth]{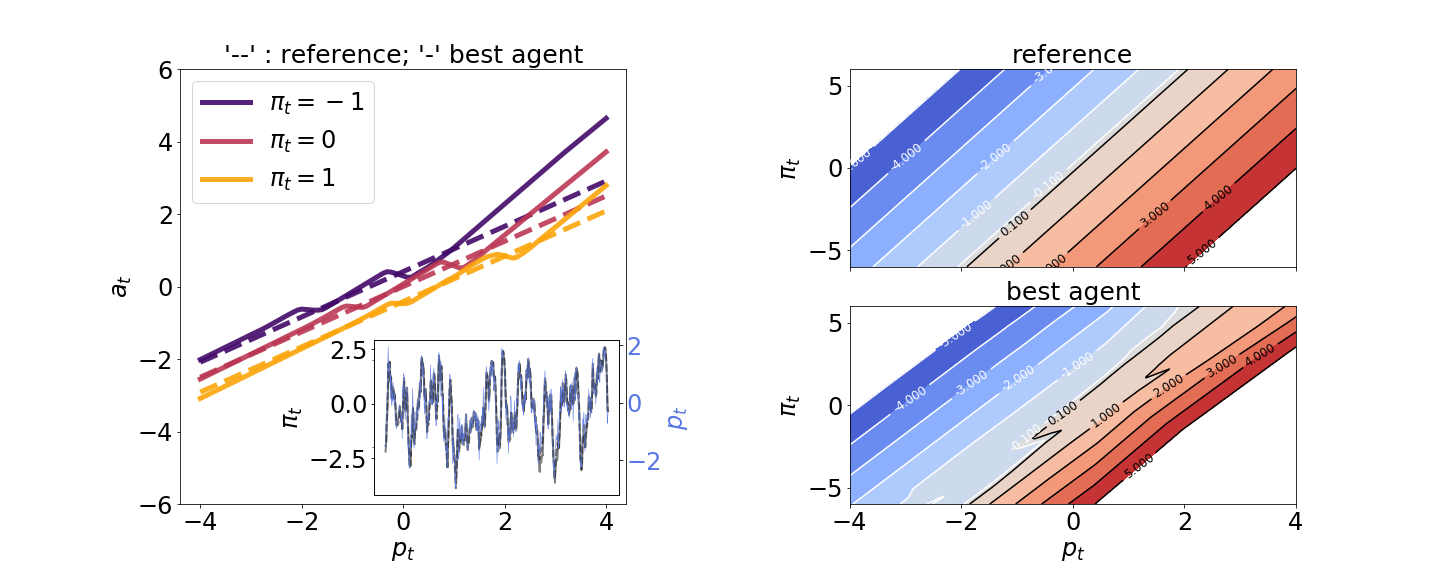}
    \caption{ Comparison of the highest-reward RL policy with the reference
      policy for the environment with quadratic cost and risk control.  Left: the actions
      $\act_t$ are plotted against the predictor $\pred_t$ for different positions
      $\pf_t\in\{-1,0,1\}$. The dashed line represent the reference strategies and the
      solid lines the RL agent. The inset compares the learnt policy with
        the reference on a small snapshot from simulated predictor's trajectories.
      Right: the contour-plots compare the reference policy (upper panel) with the RL
      policy (lower panel) in the whole $(\pf_t, \pred_t)$-plane.  }
    \label{fig:combined_policy_lqr_noiseless}
\end{figure}

\begin{figure}[h!]
    \centering
    \includegraphics[width=\textwidth]{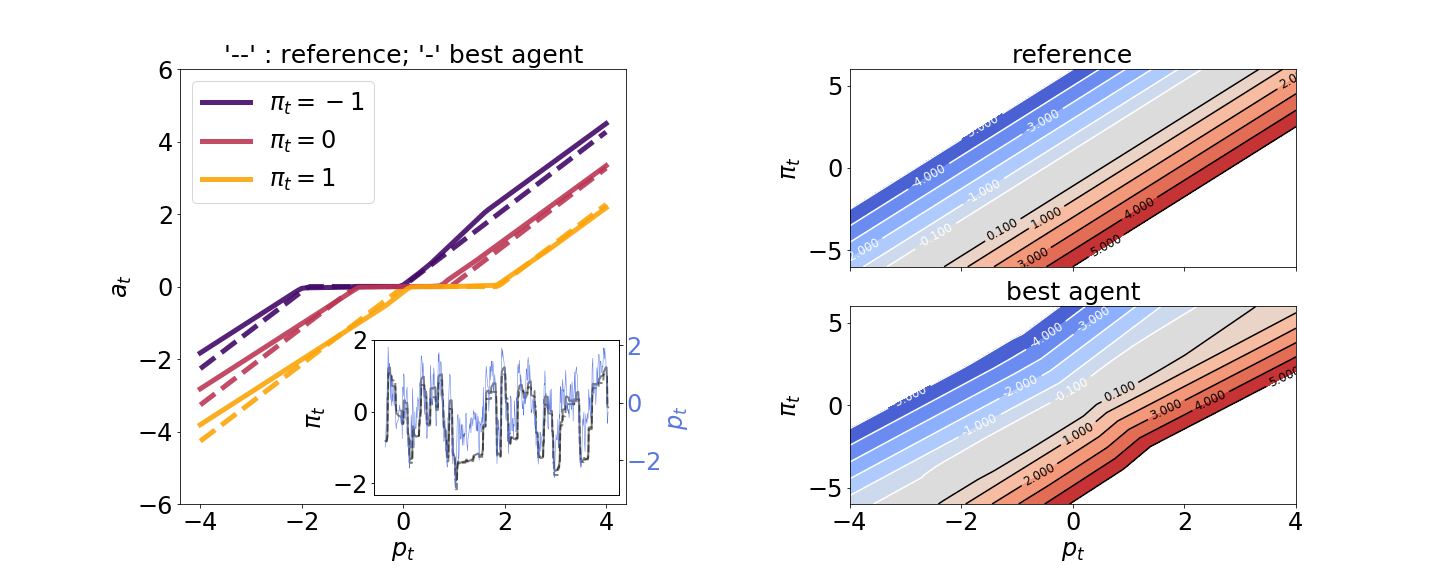}
    \caption{As Fig.~\ref{fig:combined_policy_lqr_noiseless}, but for the environment with
      linear trading cost and quadratic risk control.}
    \label{fig:combined_policy_band_noiseless}
\end{figure}

\begin{figure}[h!]
    \centering
    \includegraphics[width=\textwidth]{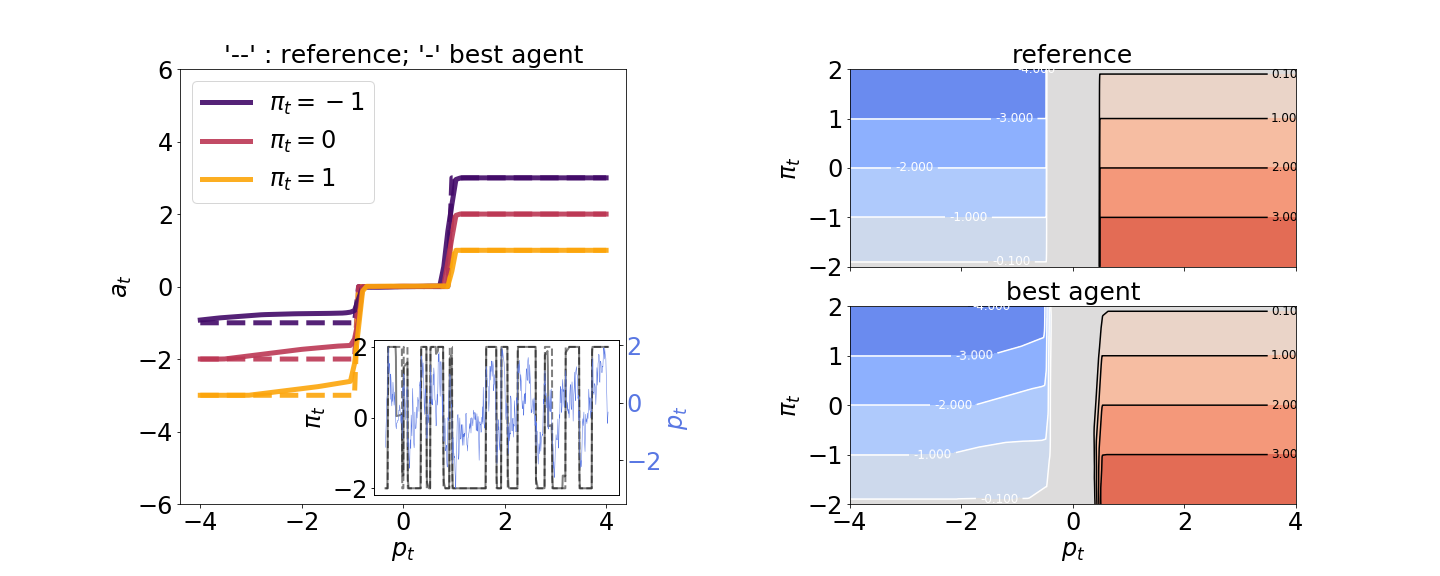}
    \caption{As Figs.~\ref{fig:combined_policy_lqr_noiseless} and
      \ref{fig:combined_policy_band_noiseless}, but for the environment with linear
      trading cost and maxpos risk control.}
    \label{fig:combined_policy_maxpos_noiseless}
\end{figure}

Besides reporting the performance in terms of the accomplished rewards it is enlightening
to take a look at the obtained policies.  In
Fig.~\ref{fig:combined_policy_lqr_noiseless}-\ref{fig:combined_policy_maxpos_noiseless} we
compare the RL policies achieving highest reward for each of the three environments with
their respective reference policies.  The figures present trading strategies in two
different ways: (a) by depicting the proposed trade $\act_t$ against the
predictor $\pred_t$ for three specific representative positions $\pf_t$. And (b) by
depicting contour plots of the proposed trade for all points in the
$(\pf_t,\pred_t)$-plane, such that $|\act_t| \leq 5$.  Similar figures for the experiments
under additional noise in the rewards can be found in~\ref{sec:appendix_noisy_rewards}.

\begin{table}
	  \caption{Summary of the performance achieved by the agents for the environment with
      quadratic trading cost and quadratic risk control.
      For $16$ agents the $75\%$-tile
      corresponds to the worst of the four best agents etc. The `Diff' calculated in the tables is the $l_1$ norm of the difference between the positions taken by the agent and those taken by the reference solution.
      }
    \centering
	\begin{tabular}{l*{6}{c}r}
		& \multicolumn{2}{c}{Reward} & \multicolumn{2}{c}{PnL} & \multicolumn{2}{c}{Diff} \\
		\hline
		\hline
		& no noise & noise & no noise & noise & no noise & noise \\
		\hline
		reference         & \multicolumn{2}{c}{0.681} &  \multicolumn{2}{c}{1.298} & \multicolumn{2}{c}{0}\\ 
		best              & 0.677 & 0.671 & 1.291 & 1.674 & 0.081 & 0.128 \\ 
		mean              & 0.665 & 0.596 & 1.237 & 1.295 & 0.140 & 0.383 \\ 
		worst             & 0.655 & 0.415 & 1.170 & 1.119 & 0.218 & 0.781 \\ 
		$75\%$-tile        & 0.668 & 0.640 & 1.258 & 1.316 & 0.161 & 0.491 \\ 
		$50\%$-tile        & 0.666 & 0.624 & 1.239 & 1.276 & 0.132 & 0.349 \\ 
		$25\%$-tile        & 0.660 & 0.588 & 1.215 & 1.215 & 0.106 & 0.256 \\ 
		\hline
		\hline
	\end{tabular}
	\label{tbl:noiseless_experiments_lqr}
\end{table}

The experiments on the environment with quadratic trading cost and risk control reliably
yield high rewards and from Tbl.~\ref{tbl:noiseless_experiments_lqr} we further draw the
conclusion, that the resulting trading strategies only show very small deviations with
respect to the reference. Nonetheless, Fig.~\ref{fig:combined_policy_lqr_noiseless} shows
that the learned policies sometimes exhibit sub-optimal features.  However, these
deviations have little negative impact in terms of the reward.  Indeed we often found
that, in the early stages of learning, the algorithm would converge to policies very close
to the reference and then become unstable and converge towards slightly sub-optimal forms.

When looking for the optimal reference strategies with a grid-search one finds that the
reward landscape w.r.t.\ the parameters is rather flat around the optimal value. Thus many
different solutions can yield almost identical results (not only for this environment).
Note also that the far ends of the policies have little influence on the resulting trading
trajectories as they are visited with small probability, permitting for larger deviations
in these regions without much negative impact on the rewards.

For the environment with linear cost and quadratic risk control the situation is rather
similar.  Clearly the RL agents retrieve the most essential features of the optimal
policy. The no-trading zones are recovered with high precision and also the slopes outside
of the no-trading zones are of an acceptable (close to linear) form. Especially in those
regions that are most explored by the predictor. Typically we find that the agents first
learn to trade into the direction of the signal and to avoid cost, before the slopes in
the trading zones are fine tuned to improve risk control.  In terms of the rewards the
solution is a little less close to the reference performances as can be gathered from
Tbl.~\ref{tbl:noiseless_experiments_band}. This should be expected for a more challenging
environment.

Fig.~\ref{fig:combined_policy_maxpos_noiseless} presents the best learnt policies for the
maxpos environment.  The agents successfully learn to take large trades once the expected
gain overcomes the cost and not to trade below this threshold. The obtained rewards
(whenever the training converged) are rather close to optimal and the situation is similar
to the previously reported environments as is further quantified in
Tbl.~\ref{tbl:noiseless_experiments_maxpos}.

\begin{table}
	\caption{Summary of the performance achieved by the agents for the environment with
      linear trading cost and quadratic risk control.}
    \centering
	\begin{tabular}{l*{6}{c}r}
		& \multicolumn{2}{c}{Reward} & \multicolumn{2}{c}{PnL} & \multicolumn{2}{c}{Diff} \\
		\hline
		\hline
		& no noise & noise & no noise & noise & no noise & noise\\
		\hline
		reference         & \multicolumn{2}{c}{0.254} &  \multicolumn{2}{c}{0.492} & \multicolumn{2}{c}{0} \\ 
		best              & 0.248 & 0.225 & 0.518 & 0.562 & 0.063 & 0.131 \\ 
		mean              & 0.241 & 0.181 & 0.478 & 0.452 & 0.093 & 0.250  \\ 
		worst             & 0.234 & 0.135 & 0.442 & 0.364 & 0.126 & 0.441  \\ 
		$75\%$-tile        & 0.244 & 0.196 & 0.491 & 0.486 & 0.101 & 0.301 \\ 
		$50\%$-tile        & 0.239 & 0.188 & 0.482 & 0.455 & 0.090 & 0.244 \\ 
		$25\%$-tile        & 0.238 & 0.167 & 0.464 & 0.414 & 0.080 & 0.181 \\ 
		\hline
		\hline
	\end{tabular}
	\label{tbl:noiseless_experiments_band}
\end{table}

The environment with maxpos risk control is the most challenging one algorithmically.  The
optimal threshold control is conceptually simple, but challenging to learn with a
model-free continuous control algorithm, such as DDPG.  We found that it was necessary to
add additional tricks for this environment in order to obtain reliable results. We
experimented with different ways to incorporate the maxpos constraint and found that the
most reliable approach is to combine a clipping of the position in the environment with an
additional soft threshold in the reward functions (cf.~algorithmic section).  One problem
is simply the shape of the function, which poses a challenge because of the non-continuity
around the non-trading region.  Another challenge is that the variance of the portfolio is
not controlled: the optimal trading strategy is to hold the position constant most of the
time and only scarcely change the position, which causes a lot of variability in the
rewards that are observed for a $(\pf_t, \pred_t)$-pair.

\begin{table}
	\caption{Summary of the performance achieved by the agents for the environment with
    linear trading cost and maxpos risk control, '-' means that some agents diverged.}
    \centering
	\begin{tabular}{l*{6}{c}r}
		& \multicolumn{2}{c}{Reward} & \multicolumn{2}{c}{PnL} & \multicolumn{2}{c}{Diff} \\
		\hline
		\hline
		& no noise & noise & no noise & noise & no noise & noise \\
		\hline
		reference         & \multicolumn{2}{c}{0.901} &  \multicolumn{2}{c}{0.901} & \multicolumn{2}{c}{0}  \\ 
		best              & 0.884 & 0.876 & 0.884 & 0.876 & 0.101 & 0.143 \\ 
		mean              & 0.856 & 0.842 & 0.856 & 0.842 & 0.198 & 0.246 \\ 
		worst             & 0.815 & 0.803 & 0.815 & 0.803 & 0.321 & 0.346 \\ 
		$75\%$-tile        & 0.849 & 0.849 & 0.849 & 0.849 & 0.148 & 0.210 \\ 
		$50\%$-tile        & 0.862 & - & 0.862 & - & 0.184 & - \\ 
		$25\%$-tile        & 0.826 & - & 0.826 & - & 0.239 & - \\ 
		\hline
		\hline
	\end{tabular}
	\label{tbl:noiseless_experiments_maxpos}
\end{table}

Overall it can be said that DDPG successfully recovers the essential features of the
reference trading strategies.  A remarkable fact is the reliable recovery of the
no-trading region of non-trivial width and trading regions of appropriate slope. This
shows that the RL agents learn all of the required features: (a) to exploit the
auto-correlation of the predictor (b) to balance risk, cost and gain appropriately.  Both
in the setting under perfect state information and under additional noise in the rewards,
the PnLs and rewards get very close to optimal. In the appendix and the tables we provide some additional
  material. Further more we provide figures, similar to
  Fig.~\ref{fig:combined_policy_lqr_noiseless}-\ref{fig:combined_policy_maxpos_noiseless}
  for all trained agents as supplemental material.

However, to obtain reliable results, the DDPG algorithm required fine adjustments,
especially in the more challenging environment with `maxpos' constraints. It is reasonable to believe that we could have obtained faster convergences and more accurate results
with more adjustments and/or better heuristics.
For example, instead of using an $\epsilon$-greedy exploration scheme, one could have considered using a parameter space noise as reported in \cite{plappert2017parameter}.

\section{Conclusions}

The aim of this paper was to demonstrate the potential of model-free reinforcement
learning methods to derive trading policies.  We first established some reference
portfolio construction problems for which we are either able to derive analytically the
trading policy or approximate it well with simple optimization procedures.  We then
compared this baseline policies with the one derived by the DDPG-RL method.

Overall it can be concluded that the DDPG-RL agents successfully recover the established
baselines. This is already a non-trivial task. Moreover, the resulting trading strategies
are very close to optimal and a high quantitative agreement with the reference strategies
have been obtained. Beyond these results, we hope that the different RL environments we built around trading optimization will be able to serve as reference test-cases in further studies.

Finally, one can notice that a specificity of the model-free RL
approach we used is that the structure of the reward function is assumed unknown.
However, for the studied problems the differentiable structure of the reward function
could have been exploited to design more efficient algorithms.
Such a model-based approach could be studied in future work.

\section*{Acknowledgements}
This research was partly conducted within the \emph{Econophysics \& Complex Systems}
Research Chair, under the aegis of the Fondation du Risque, the Fondation de l'Ecole
polytechnique, the Ecole polytechnique and Capital Fund Management.

We would like to thank Nathanaël Foy, Alexis Saïr and Julien Lafaye for regular
discussions on the content of this article, and Charles-Albert Lehalle for sharing his expertise on the topic.

\section*{Conflict of interest}
The authors declare that they have no known competing financial interests or personal relationships that could have appeared to influence the work reported in this paper.

\bibliographystyle{elsarticle-num-names}

\newpage

\appendix

\section{Results under additional noise in the rewards}
\label{sec:appendix_noisy_rewards}

Here we present the results when the observed rewards contain an additional noise in
the returns, instead of the actual values of the predictor. The resulting signal has a
noise-to-signal-ratio of $10$ for the cases with quadratic risk control, and 
$4$ for the maxpos risk control since its convergence is less stable.

\begin{figure}[h!]
    \centering
    \includegraphics[width=\textwidth]{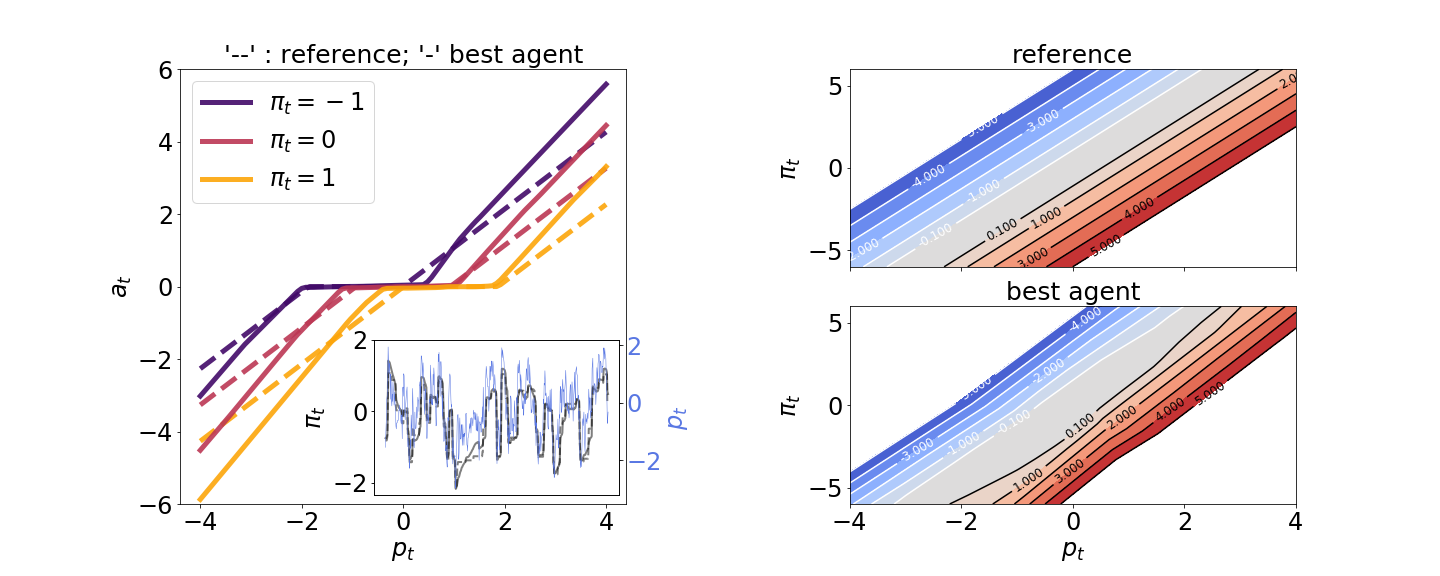}
    \label{fig:combined_policy_lqr_noise}
    \caption{Highest-reward RL agent for the environment with quadratic trading
      cost and quadratic risk control. As Fig.~\ref{fig:combined_policy_lqr_noiseless},
      but with additional noise in the rewards.}
\end{figure}

\begin{figure}[h!] \centering
\includegraphics[width=\textwidth]{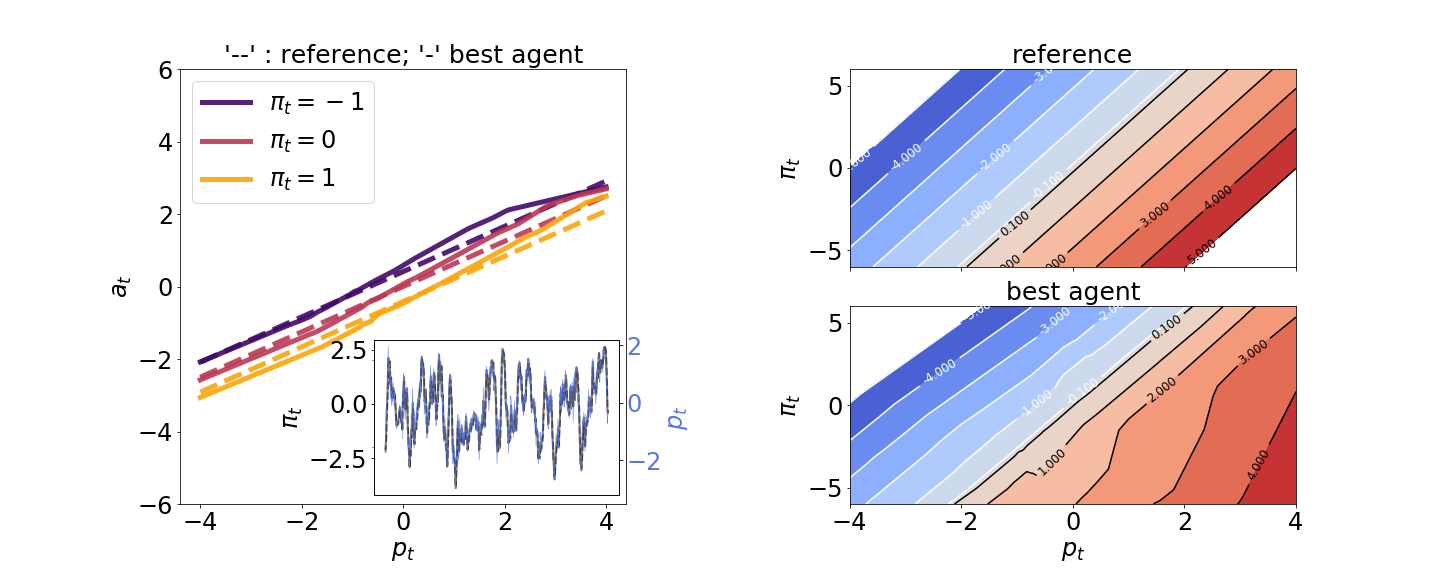}
    \label{fig:combined_policy_lin_trade_noise}
    \caption{Highest-reward RL agent for the environment with proportional
trading cost and quadratic risk control. As Fig.~\ref{fig:combined_policy_band_noiseless},
but with additional noise in the rewards.}
  \end{figure}

\begin{figure}[h!] \centering
\includegraphics[width=\textwidth]{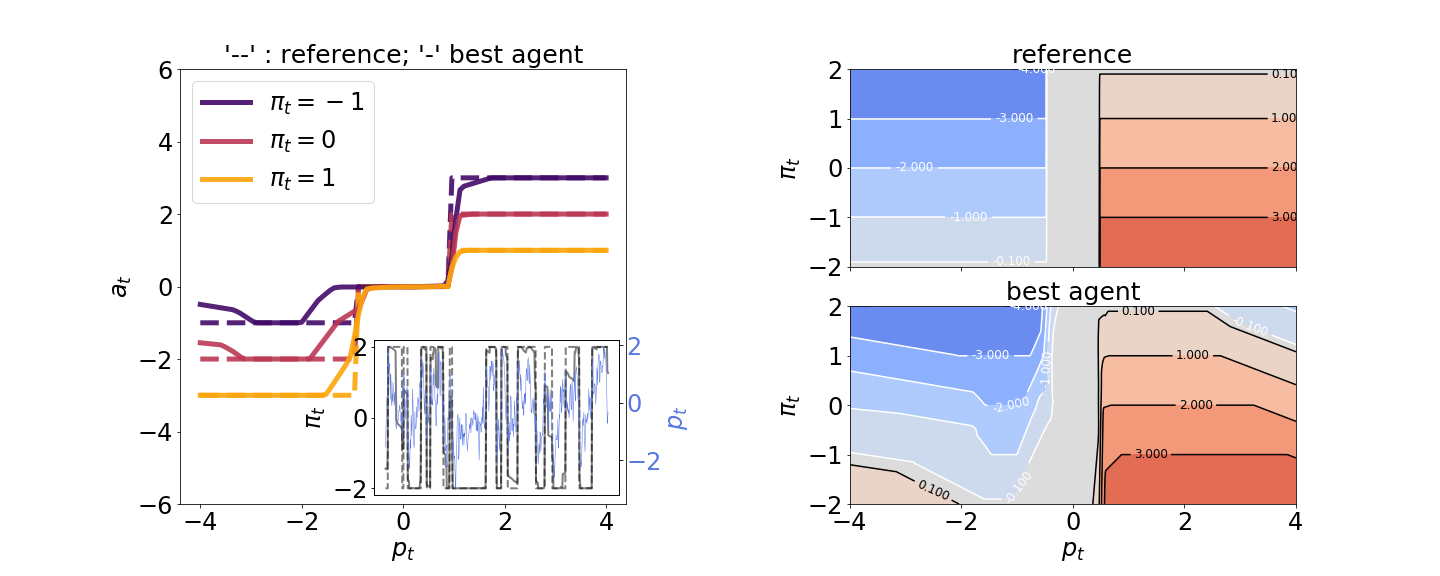}
    \label{fig:combined_policy_maxpos_noise}
    \caption{Highest-reward RL agent for the environment with proportional
trading cost and maxpos risk control. As Fig.~\ref{fig:combined_policy_maxpos_noiseless}, but with additional noise in the rewards.}
\end{figure}

\section{Details of the algorithm}
\label{sec:appendix_algo_details}

Let us briefly recall the basic elements of DDPG. DDPG is an actor-critic architecture
where the actor estimates a deterministic policy while the critic estimates the
state-action value function $\qf$. This algorithm is training off-policy relying on the
off-policy deterministic policy gradient theorem \cite{silver2014deterministic}.

Formally, let $\beta\left(\act|\state\right)$ denote a behaviour policy that generates the
training samples and $\rho^\beta$ be the state distribution under $\beta$. The objective
function in the off-policy setting as defined in \cite{degris2012off} is
\begin{align*}
    \mathcal{J}^{\beta}\left(\params \right) &= \int_{\state\in \mathcal{S}}\rho^{\beta}\left(\state \right)\vf_{\params}\left( \state\right)d\state \\
    &= \int_{\state \in \mathcal{S}}\rho^{\beta}\left( \state\right)\qf_\params\left(\state, \policy_\params\left(\state \right) \right)d\state
    \, .
\end{align*}
Here $\policy_\params$ stands for our parametric deterministic policy with parameter
$\params$, $\mathcal{S}$ the state space and $\qf_\params = \qf^{\policy_\params}$ the
state-action value function of policy $\policy_\params$.

The off-policy policy gradient theorem states:
\begin{equation*}
    \nabla_\params \mathcal{J}^\beta\left( \params\right) \approx \mathop{\ee}_{\state \sim \rho^\beta}
    \left[ \nabla_\params \policy_\params\left( \state\right)\nabla_\act\qf_\params\left(\state, \act \right)|_{\act=\policy_\params\left( \state\right)}\right].
\end{equation*}
This estimation is useful as it can be used to update $\params$ with gradient ascent,
without having to take into account the dependency of $\qf_\params$ on $\params$.

\begin{figure}    \centering
    \includegraphics[scale = 0.3]{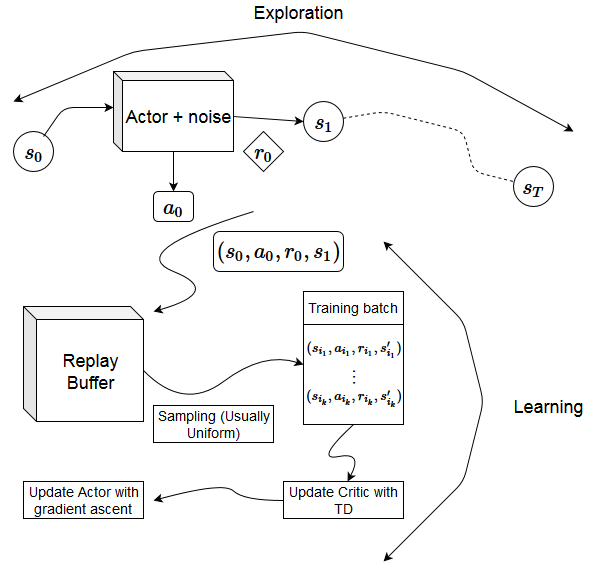}
    \caption{Off-policy training summary.}
    \label{fig:Actor-Critic}
\end{figure}

In off-policy training the agent interacts with the environment to gather experiences
(training samples in the form of tuples
$\left( \state_t, \act_t, \reward_t, \state_{t+1}\right)$ of state, action, reward, next
state) in a replay buffer, then samples training batches in order to update first the
critic and then the actor (see the Fig.~\ref{fig:Actor-Critic}).

\subsubsection*{Exploration}

An episode represents a walk through the environment, the agent always starts with a
position $\pf_0=0$. The agent interacts with the environment by taking actions
$\act_t = \act_t^{pred} + \anoise_t$, given the current position $\pf_t$ and the predictor
$\pred_t$:
\begin{align*}
\begin{cases}
\act^{pred}_t &= \policy^{\params}\left(\pred_t, \pf_{t} \right) \\
\anoise_{t} &= (1-\rho^{\mathrm{expl}}) \anoise_{t-1} + \sigma^{\mathrm{expl}} \epsilon_t; \\
\anoise_0 &= 0
\end{cases}
\, .
\end{align*}
The $\anoise_t$ is the exploration noise, following an autoregressive process, with
i.i.d.~$\epsilon_t$ s.t.~$\epsilon_t \sim \mathcal{N}\left(0, 1\right),
\sigma^{\mathrm{expl}} > 0$.

\subsubsection*{Learning}

Every $\tau$ steps we sample a training batch $\left(\state_{t_j}, \act_{t_j},
\reward_{t_j}, \state_{t_j+1} \right)_{1 \le j \le b}$ of size $b$ from the replay
buffer. This decorrelates experiences within a batch so that gradient estimation is less
biased.

\paragraph*{Critic update}

Let $\qf$, $\widetilde{\qf}$, $\policy$ and $\widetilde{\policy}$ denote respectively the
current critic, target critic, current actor and target actor networks with their
respective parameters $w, \widetilde{w}, \params, \widetilde{\params}$.  For a given
experience $i$ in the sampled training batch, the target $\qf$-value is computed with the
target critic network where the action is chosen with the target actor network
$\widetilde{\policy}$
\begin{align*}
\widetilde{Q}_i = \reward_{t_i} + \gamma \widetilde{\qf}\left(\state_{t_i+1}, \widetilde{\policy}\left(\state_{t_i+1} \right) \right)
\end{align*}
And the $\qf$-value is simply computed with the current critic network
\begin{align*}
Q_i = \qf\left(\state_{t_i}, \act_{t_i} \right)
\end{align*}
Then the critic loss is calculated with temporal difference:
\begin{align*}
\mathcal{L}_{critic}\left(w \right) = \frac{1}{2b}\sum_{i=1}^b\left(\widetilde{Q}_i - Q_i \right)^2
\end{align*}
Where $b > 0$ is the batch size.

\paragraph*{Actor update}

Using the off-policy deterministic policy gradient theorem the current actor network is
updated with gradient descent on the following loss function:
\begin{align*}
\mathcal{L}_{actor}\left(\params \right) &= -\frac{1}{b}\sum_{i=1}^b\qf\left(\state_i, \policy_\Theta \left( \state_i\right) \right)
\, .
\end{align*}

\paragraph*{Target networks update}

The parameters of target Critic and target Actor networks are updated with soft target updates
\begin{align*}
    \Tilde{w} &\leftarrow \tau_{critic}w + \left(1 - \tau_{critic} \right)\Tilde{w} \\
    \widetilde{\params} &\leftarrow \tau_{actor}\params + \left(1 - \tau_{actor} \right)\widetilde{\params}
\end{align*}
Where $0 < \tau_{critic} < 1$ and $0 < \tau_{actor} < 1$, with this method, the networks
parameters get updated slowly, which makes training more stable.

\paragraph*{The replay buffer}

Different schemes may be applied when sampling from the replay buffer. We choose
prioritized experience replay \cite{schaul2015prioritized} because it can speed up
training and improve convergence. In prioritized experience replay buffers, experiences
are weighted according to their TD error
\begin{align*}
\delta\left(\state, \act, \reward, \state' \right) = \reward + \gamma \max_{\act'}\qf\left(\state', \act' \right) - \qf\left(\state, \act \right)
\end{align*}
and are sampled according to the distribution
\begin{align*}
P\left( i\right) = \frac{p_i^\alpha}{\sum_k p_k^\alpha}, \hspace{2ex}
1 \le i \le N, \hspace{2ex} 0 \le \alpha \le 1
\end{align*}
Where $N$ is the replay buffer size, $p_i = |\delta_i| + \epsilon$ the priority with
$\epsilon$ is a small positive number ensuring non-zero sampling probability for all
samples.

The intuition is that experiences with high magnitude TD errors are those poorly evaluated
by the critic, so increasing their learning frequency makes sense.

Notice that non-uniform sampling introduces a bias in the gradient estimation that can be
corrected with importance sampling weights \cite{mahmood2014weighted} by using
$\alpha_i\delta_i$ instead of $\delta_i$ in updating the critic, with
\begin{align*}
\alpha_i = \left(\frac{1}{N}\frac{1}{P\left(i \right)} \right)^\beta,\hspace{2mm} 0 \le \beta \le 1
\, .
\end{align*}
The exponent $\beta$ represents the emphasis put into correcting the bias. In our
experiments this parameter is linearly annealed from an initial value $\beta_0$ to unity
as correcting the bias is not as important in the beginning as it is near convergence.

\end{document}